\documentclass[11pt]{amsart}
\usepackage{amsfonts,eucal}
\begin{document}

\newcommand{\bx}{{\bf x}}
\newcommand{\by}{{\bf y}}
\newcommand{\bM}{{\overline{\mathcal M}}}
\newcommand{\half}{{\textstyle{\frac{1}{2}}}}
\newcommand{\quarter}{{\textstyle{\frac{1}{4}}}}
\newcommand{\R}{{\mathbb R}}
\newcommand{\bP}{{\mathbb P}}

\title{Notes on two elementary evolutionary games}
\author{Jack Morava}
\address{The Johns Hopkins University,
Baltimore, Maryland 21218}
\email{jack@math.jhu.edu}
\subjclass{{\bf draft}}
\date {31 October 2010}

\begin{abstract}{
Thus spoke the wise Queen Delores, saying, ``I have studied mathematics. I will question this
young man, in my tent tonight, and in the morning I will report the truth as to his
pretensions''.\\
{}\\
{\bf Jurgen}, James Branch Cabell (1919), Ch. XXXII: Jurgen proves 
it by mathematics}\end{abstract}

\maketitle

\noindent Evolutionary game theory [3] defines something like a functor from the classical theory
of games to dynamical systems, imagining biological entities whose rate of reproduction is
proportional to their success at playing the game in question. It is a beautiful and accessible
subject; these notes on two interesting examples grew out of JHU's 2007 BioCalc I course, and
I'd like to thank the students there for their interest and forebearance. This work was also
suggested by DARPA's Fundamental Problems of Biology initiative. \bigskip

\section{The Battle of the Sexes} \bigskip

\begin{quotation}{'' C'est magnifique, mais ce n'est pas la guerre \dots'' \medskip

\noindent
Marshal Pierre Bosquet, Balaclava 1854}\end{quotation} \bigskip

\noindent
{\bf 1.1} In the evolutionary version of Richard Dawkins' toy model for marriage markets as
asymmetric two-player games, the rate of reproduction of a population type is proportional to its
success at beating the mean expectation for the game. Following Hofbauer and Sigmund [3 \S
10.2], this is expressed by a system

\[
\dot{x_k} \; = \; x_k \cdot [(A\by)_k - \bx \cdot A \by] \;,
\]
\[
\dot{y_k} \; = \; y_k \cdot [(B\bx)_k - \by \cdot B \bx] \;.
\]

\noindent
of `replicator' equations, in which there are two types of players: those of type I (males) 
characterized by a state vector $\bx = (x_1,\dots,x_n)$ in a unit simplex in some space of
strategies, and type II (females) defined similarly by a vector $\by = (y_1,\dots,y_m)$. \bigskip

\noindent
The details of the model are specified by matrices $A,B$: a type I player choosing mixed strategy
$\bx$ against  a type II player choosing strategy $\by$ receives payoff $\bx \cdot A \by$, while
his type II `opponent' [the game is {\bf not} zero-sum!] receives payoff $\by \cdot B \bx$. In
the simplest version of the game both types of players have two strategies:
\[
\bx \; = \; (x, 1 - x), \; \by \; = \; (y, 1 - y) \;.
\]
The traditional terminology is that $x(t)$ is the proportion of `fast' males and $y(t)$ is the
proportion of `slow' females; this choice of parameters foregrounds analogies with a
Lotka-Volterra system. \bigskip

\noindent
The payoff matrices are
\[
A = 
\left[\begin{array}{cc}
                         0               &               G \\
                G - E - \half C   & G - \half C \\
\end{array}\right]
\]
and
\[
B = 
\left[\begin{array}{cc}
                  0     &    G - E - \half C \\
             G - C   &          G - \half C \\
\end{array}\right] \;.
\]
In the four possible types of interaction, a fast (`macho'?) male in an encounter with fast female
receives payoff $G$ while the female receives $G - C$; when a fast male encounters a slow
(`coy'?) female both receive payoff 0. When a slow male encounters a fast female, both players
receive payoff $G - \half C$, and in an encounter between a slow male and a slow female both
players receive payoff $G - E - \half C$. The model assumes that
\[
 0 \; < \; E \; < \; G \; < \; C < \;2(G - E) \;;
\]
the engagement cost $E$ (borne equally by the players) is less than the (individual) payoff $G$
for successful reproduction, and the total cost $C$ of reproduction, though bigger than the payoff
for a single participant, is less than the total payoff to both parties, less the total cost of
engagement. \bigskip  

\noindent
{\bf 1.2} The resulting system 
\[
\dot{x} \; = \; x (1 - x) [ \half C  + (E - G) y] 
\]
\[
\dot{y} \; = \; y ( 1- y) [ - E +  (C + E - G) x]
\]
of replicator equations is completely integrable, with five critical points, hyperbolic at the corners
of the unit square\begin{footnote}{Corresponding to classical models: the Garden of Eden, the
Summer of Love, Total War, and Christian Heaven \dots}\end{footnote} and a more interesting 
elliptic point at 
\[
X \; = \; \frac{E}{E-G+C} \; , \;  Y \; = \; \frac{C}{2(G - E)} \;.
\]
The linearization near this critical point is defined by the Hessian or Jacobian matrix of the 
right-hand side of this system; it has purely imaginary eigenvalues $\Lambda$ satisfying
\[
\Lambda^2 =  - \frac{C E (2(G - E) - C)(C - G)}{4(G - E)(C + E - G)} \;.
\]
Any point in the interior of the square lies on a closed orbit, whose period approaches 
$T = 2 \pi |\Lambda|^{-1}$ at the fixed point. The desire to understand this relatively
complicated `observable' was the initial motivation for this note. \bigskip
                         
\noindent
{\bf 1.3 Proposition} $(E,G,C) \mapsto (X,Y)$ extends to a map
\[
[E:G:C] \mapsto [EC:2E(G - E):C(E - G + C):2(G - E)(E - G + C)] 
\]
from $\bP_2$ to a quadric surface $\bP_1 \times \bP_1 \subset \bP_3$. \medskip

\noindent
{\bf Proof:} If we write $[Z_0:Z_1:Z_2:Z_3]$ for the coordinates in $\bP_3$, then evidently $X =
Z_0Z_2^{-1}, \; Y = Z_1 Z_3^{-1}$; in other words, the map above is the composition of the
map from Dawkins' space of economic parameters $E,G,C$ to psychosocial parameters $X,Y$
followed by the Segre embedding
\[
(x_0,x_1) \times (y_0,y_1) \mapsto (x_0y_0, x_0y_1, x_1y_0,x_1y_1) 
\]
of the quadric surface $Z_0Z_3 = Z_1 Z_2$; for example, $[1:1:0] \mapsto [1:0:0:0]$ and
$[0:1:1] \mapsto [1:1:0:1]$. $\Box$ \bigskip

\noindent 
The function $\Lambda$ is homogeneous of weight one in the economic parameters; its value
thus depends on a choice of units. This issue is familiar from physics: similar considerations in
the case of the van der Waals model for liquid-gas transitions, for example, led historically to
thermodynamics' law of corresponding states [2 \S 6.3]. \bigskip

\noindent
{\bf Corollary} 
\[
\Lambda^2 \; = \; (2Z_3)^{-1}(Z_0 - Z_1) 
\]
in units defined by the geometric mean of $C$ and $C - G$. \medskip

\noindent
[More precisely, we have 
\[
\Lambda^2 \; = \; \frac{Z_0 - Z_1}{2Z_3} \cdot C(C-G) \;, 
\]
and in `natural' units such that $C(C-G) =1$, ie 
\[
G = C - C^{-1}
\]
we can omit the factor on the right.] \bigskip

\section{Progressives and Conservatives}\bigskip

\begin{quotation}{``If I can't sell it gonna sit down on it,\\
never catch me givin it away!''\medskip

\noindent
Ruth Brown, Fantasy Records (1989)}\end{quotation} \bigskip

\noindent
{\bf 2.1} Similar techniques can be used to analyze the much simpler game defined by payoff
matrices
\[
A = a H, B = bH^{\rm T}
\]
with 
\[
H  = 
\left[\begin{array}{cc}
                         0        &    \eta((b-a) \\
                    \eta(a-b)  &    0\\
\end{array}\right] \;,
\]
where $\eta(x)$ is the Heaviside function ($=1$ if $x > 0, \; = \half$ if $x=0$, and $=0$
otherwise; I'll assume that $a$ and $b$ are both positive, and that $a \neq b$ to exclude trivial
cases.  In this example there are again two types of players, with strategy vectors $(x,1-x)$ and
$(y,1-y)$ as above, now interpreted as the proportion of progressive (resp. conservative)
participants of type I (resp II); with payoff parameters $a,b$ for the two types. \bigskip

\noindent
An encounter between a progressive and a conservative yields zero, unless the conservative
receives the larger payoff. An encounter between two progressives yields payoff $a$ for the type
I player and payoff $b$ for the type II player, and an encounter between two conservatives
produces nothing for either. This is therefore a rather silly game: the progressive agrees with
anything that benefits anybody, while the conservative strategy amounts to pure bullying; but
because the game is not zero-sum, it is not completely trivial. \bigskip

\noindent
{\bf 2.2} The obvious symmetries of the payoff matrices result in very simple replicator
equations

\[
\dot{x} \; = \; a x (1 - x) \; \eta(b - a)
\]
\[
\dot{y} \; = \; b y (1 - y) \; \eta(a - b) \;.
\]

\noindent
One or the other of these equations is thus trivial, depending on the relative sizes of $a$ and $b$:
the proportion of participants whose payoff is larger (let's call them fat cats) does not change
with time, while the group with lower payoff (underdogs?) become more progressive, following
a logistic growth pattern; the population thus evolves toward political polarization. \bigskip

\bibliographystyle{amsplain}

\end{document}